\documentclass[12p]{iopart} 
\usepackage{hyperref}
\usepackage{graphics} 
\usepackage{graphicx} 
\usepackage{subfigure}
\usepackage{url}
\usepackage{tipa}
\usepackage{iopams}
\usepackage{braket}
\usepackage{cite}

\newcommand{\field}[1]{\mathbb{#1}} 

\begin{document} 
\article{}{Self-interference of a toroidal Bose-Einstein condensate}
\author{L A Toikka}
\address{Turku Centre for Quantum Physics, Department of Physics and Astronomy, University of Turku, 20014 Turku, Finland}
\ead{\mailto{laantoi@utu.fi}}

\begin{abstract}
We demonstrate the role played by a ring trap and the associated ring geometry in inducing self-interference of a toroidal Bose-Einstein condensate. We start by showing how the existence of the self-interference can be inferred from the Wigner function representation of the ring. Then, using analytical expressions for the time-evolution of a freely expanding ring condensate with and without a persistent current, we show that the self-interference of the ring condensate is possible only in the absence of the persistent current. We conclude by proposing an experimental protocol for the creation of ring dark solitons using the toroidal self-interference.
\end{abstract}
\pacs{03.75.Kk, 03.75.Lm}

\maketitle

\section{\label{sec:Introduction}Introduction}

Following the pioneering demonstration of interference between two freely expanding Bose-Einstein condensates (BECs)~\cite{Andrews31011997}, also self-interference of a single condensate has been experimentally and numerically observed in hard-wall reflections~\cite{PhysRevLett.83.3577,PhysRevA.63.043613}. Later generalisations to optical lattices~\cite{Bloch05} have led to interference between as many as 30 separate BECs~\cite{Anderson27111998,PhysRevLett.93.180403}. 

In a self-interference setting, the wavefunction splits up into two (or more) pieces, which travel different paths; subsequently the self-interference arises when both parts are later spatially recombined. This process can happen next to a hard-wall potential, as mentioned, whereby the condensate interferes with its reflection. However, a natural setting to consider the self-interference of a BEC in the absence of any boundary effects is a non-simply connected geometry such as a torus. A BEC in a ring trap forms an extended quantum object that winds back to itself, and by controlling the size of the central forbidden region, one might expect the separate parts of the same condensate to show self-interference phenomena.

As the phase itself is not an observable, the appearance of interference fringes reflects the relative long-range coherence between the separate condensates, or different parts of the same condensate. Even in the case of well-defined atom numbers and fully uncertain phases, interference is produced by the action of the quantum measurement, which assigns instantaneous relative phase correlations~\cite{PhysRevLett.76.161}. In the case of self-interference, this action is equivalent to the condensate encountering itself. The phase coherence of BECs also leads to many other important properties such as quantised vortices~\cite{PhysRevLett.98.110402} and dark solitons~\cite{Greekreview}.

In general, ring traps~\cite{PhysRevLett.80.3899,PhysRevLett.78.4713,Griffin2008,Baker2009,Sherlock2011,PhysRevA.74.023617,PhysRevLett.99.083001,PhysRevA.63.013608,1367-2630-10-4-043012} have attracted significant interest in the form of e.g. persistent currents~\cite{PhysRevA.58.580,PhysRevA.86.011602}, atomic-phase interference devices~\cite{PhysRevA.67.033601}, vortex dynamics~\cite{PhysRevA.64.063602,PhysRevA.77.032107}, and ring dark solitons~\cite{PhysRevE.49.1657,ths2012,PhysRevA.87.043601,2013arXiv1309.0732T}, but their role in inducing self-interference is an open question. Also, given the high experimental relevance of ring traps, observation of toroidal self-interference offers one way to probe the spatial phase coherence over the extent of the condensate in the ring, and it can also serve as a mechanism for the creation of ring dark solitons, as we show in~\Sref{sec:RDS}.

In the work presented here, we investigate in detail how the self-interference of a Bose-Einstein condenstate can appear in a ring trap. We consider analytically the Wigner function representation of a Gaussian ring condensate, showing that if the radius of the ring trap is brought to zero, all the self-interference phenomena disappear, as expected. For non-zero radii, on the other hand, we demonstrate a possible experimental way to observe the self-interference by letting the ring expand freely and overlap with itself to produce circular fringes. Interestingly, we find that in the presence of a persistent current, there cannot be any self-interference because the central vortex is enforcing a density hole at $r = 0$, preventing the condensate from overlapping with itself. Insofar as the centrifugal potential barrier acts as a circular hard wall, we will not consider the separate case of self-interference arising from reflection of the condensate. Finally, based on the toroidal self-interference, we propose a protocol for the experimental creation of ring dark solitons.

\section{\label{sec:tb}Theoretical background}

We consider a scalar order parameter $\psi$, representing the macroscopic wavefunction of a Bose-Einstein condensate trapped in a potential given by $V_{\mathrm{trap}}$, which is a solution to the Gross-Pitaevskii equation (GPE):
\begin{equation}
\label{eqn:nls0}
\rmi \psi_t = -\nabla^2 \psi + V_{\mathrm{trap}}\psi + C_{\mathrm{2D}}|\psi^2|\psi.
\end{equation}
Here we have assumed a two-dimensional condensate whereby the $z$-direction is tightly trapped to the corresponding harmonic oscillator ground state ($\omega_z \gg \omega_x = \omega_y \equiv \omega_r$) and has been projected onto the $xy$-plane. Then $C_{\mathrm{2D}} = 4 \sqrt{\pi} Na/a_{\mathrm{osc}}^{(z)}$, where $N$, $a$, and $a_{\mathrm{osc}}^{(z)}$ are the number of atoms in the cloud, the $s$-wave scattering length of the atoms, and the characteristic trap length in the $z$-direction respectively. We have obtained dimensionless quantities by measuring time, length and energy in terms of $\omega_r^{-1}$, $a_{\mathrm{osc}}^{(r)} \equiv a_{\mathrm{osc}} = \sqrt{\hbar/(2m\omega_r})$ and $\hbar \omega_r$ respectively, where $\omega_r$ is the angular frequency of the trap in the $r$-direction. This basis is equivalent to setting $\omega_r = \hbar = 2m = 1$. All the units in this work are expressed in this dimensionless basis. Using the experimental parameters of the toroidal BEC of~\cite{PhysRevLett.111.205301} and $^{87}\mathrm{Rb}$, we obtain $a_{\mathrm{osc}} = 0.8\, \mu\mathrm{ m}$, $C_{\mathrm{2D}} \sim 200-2000$, and one unit of time is $\sim 1.5$ ms.

We consider the toroidal condensate to be in the ground state of the harmonic ring trap given by 
\begin{equation}
\label{eqn:Vtrap}
V_{\mathrm{trap}} = \frac{1}{4}\omega^2(r-r_0)^2,
\end{equation}
where we take $r_0 = 4.5$, and $\omega = 20$. To a very good approximation, the ground state of the potential~\eref{eqn:Vtrap} with a weak nonlinear interaction of $C_{\mathrm{2D}} = 50$ is given by a Gaussian, 
\begin{equation}
\label{eqn:GaussApprox0}
\varphi_G (r) = \mathcal{N} \rme^{-\frac{(r-r_0)^2}{2\sigma^2}},
\end{equation}
where $\mathcal{N}$ is the normalisation, and $\sigma = \sqrt{2/\omega}$.

\section{Self-interference in the torus}
\subsection{\label{sec:Wigner}Wigner function of the ring}

The effect of the ring trap in inducing self-interference can be seen in the Wigner function~\cite{PhysRev.40.749, Hillery1984121}. There exists a one-to-one mapping between the vector $\Braket{\textbf{r}|\psi} = \psi(\textbf{r})$ and the corresponding Wigner function $W_\psi (\textbf{r},\textbf{p}) \in \field{R}$, defined for our two-dimensional states by
\begin{equation}
\label{eqn:WignerF}
W_\psi (\textbf{r},\textbf{p}) = \frac{1}{(2\pi)^2}\int_{\field{R}^2}\mathrm{d}\textbf{u}\, \psi^* \left(\textbf{r}-\frac{\textbf{u}}{2}\right) \psi \left(\textbf{r}+\frac{\textbf{u}}{2}\right) \rme^{-\rmi \textbf{u} \cdot \textbf{p}}.
\end{equation}
The Wigner function can be thought of as being another name or representation for the state $\Ket{\psi}$, containing all the possible information of the state, with the important property that the marginals of the Wigner function recover the position and momentum distributions:
\numparts
\begin{eqnarray}
\label{eqn:Wigner_marg_r}
\int_{\field{R}^2} \mathrm{d}\textbf{p}\, W_\psi (\textbf{r},\textbf{p}) &= |\Braket{\textbf{r}|\psi}|^2, \\
\label{eqn:Wigner_marg_p}
\int_{\field{R}^2} \mathrm{d}\textbf{r}\, W_\psi (\textbf{r},\textbf{p}) &= |\Braket{\textbf{p}|\psi}|^2.
\end{eqnarray}
\endnumparts
A necessary and sufficient condition for the Wigner function to be a true phase space density~\cite{Hudson1974249} states that this happens only for coherent and squeezed vacuum states (Gaussian states). However, the Wigner function representation is often quite suitable for presenting nonclassical states, and in our case, the negativity of the Wigner function is associated with the emergence of toroidal matter-wave interference.

\begin{figure}
\centering
\subfigure{\includegraphics[width=0.45\textwidth] {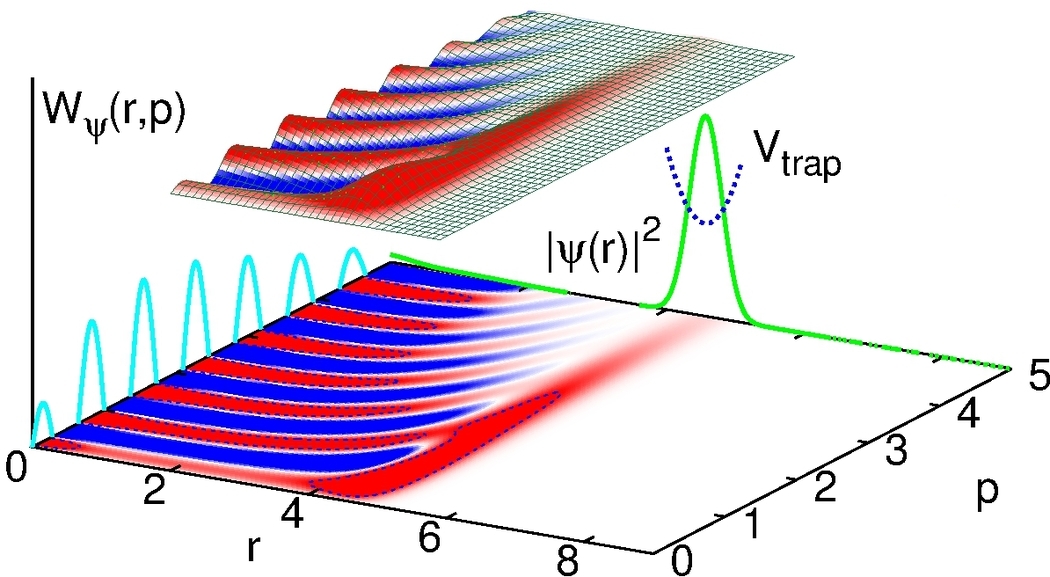}} 
\subfigure{\includegraphics[width=0.35\textwidth] {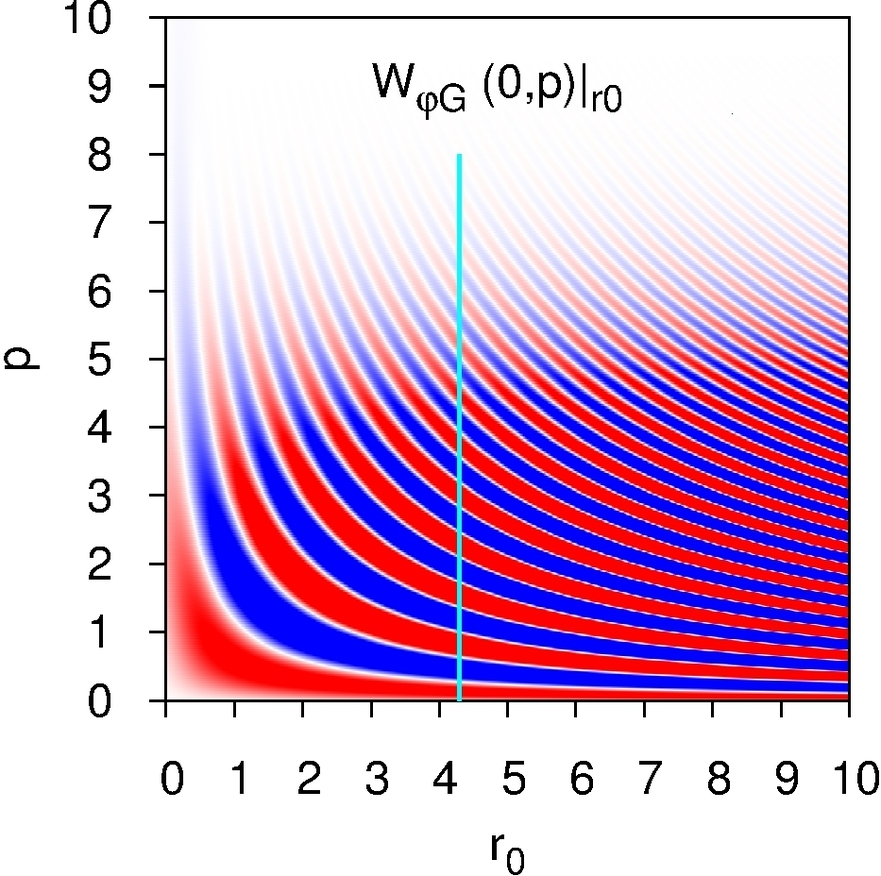}} \\
\caption{\label{fig:GaussApprox1} Left panel: The Wigner function $W_{\psi}(r,p)$ of the ground state of the potential~\eref{eqn:Vtrap} with $r_0 = 4.5$. Taking the marginal over $p$ recovers the original spatial density (solid green line). Right panel: The Wigner function $W_{\varphi_G}(0,p)|_{r_0}$ [see~\Eref{eqn:GaussApprox1}] of a toroidal condensate of radius $r_0$ at the centre of the ring ($r = 0$). Self-interference fringes arising from the toroidal geometry appear for $r_0 > 0$. The light blue line shows the case corresponding to the left panel (within numerical accuracy). Red (blue) color corresponds to positive (negative) values. }
\end{figure}
Substituting~\Eref{eqn:GaussApprox0} into~\Eref{eqn:WignerF}, and setting $r = 0$ as we are interested in the neighbourhood of the origin, we obtain
\begin{equation}
\label{eqn:GaussApprox1}
W_{\varphi_G}(0,p) = \mathcal{N}^2 \int_{0}^{\infty} \mathrm{d} u\, u p J_0 (up) \rme^{-\frac{(u/2-r_0)^2}{\sigma^2}},
\end{equation}
where $J_0$ is the Bessel function of order 0. We can already see that $W_{\varphi_G}(0,p)|_{r_0 = 0} = 2\mathcal{N}^2 p\sigma^2 \rme^{-p^2 \sigma^2}$, which is manifestly positive, i.e. there are no fringes. The interference effects due to the toroidal geometry vanish when the radius of the torus is taken to zero. To consider the case of $r_0 > 0$, the integral in~\Eref{eqn:GaussApprox1} can be evaluated analytically with a small correction that must be calculated numerically (see the Appendix).

The interference effects induced by the toroidal geometry (of radius $r_0$) are shown in Fig.~\ref{fig:GaussApprox1} in terms of the Wigner function. Considering the case of $r_0 = 4.5$ (left panel), the fringes in the middle of the torus show that the opposite parts of the ring are interfering with each other. As expected, for $r_0 = 0$ there can be no toroidal self-interference, and the Wigner function is manifestly positive (right panel). As $r_0$ is being increased from zero and therefore the ring gets a non-zero radius, we can see the appearance of interference fringes in $p$ (at the origin of the ring $r = 0$), with the fringe separation decreasing as $r_0$ increases. 

\begin{figure}
\centering
\includegraphics[width=\textwidth] {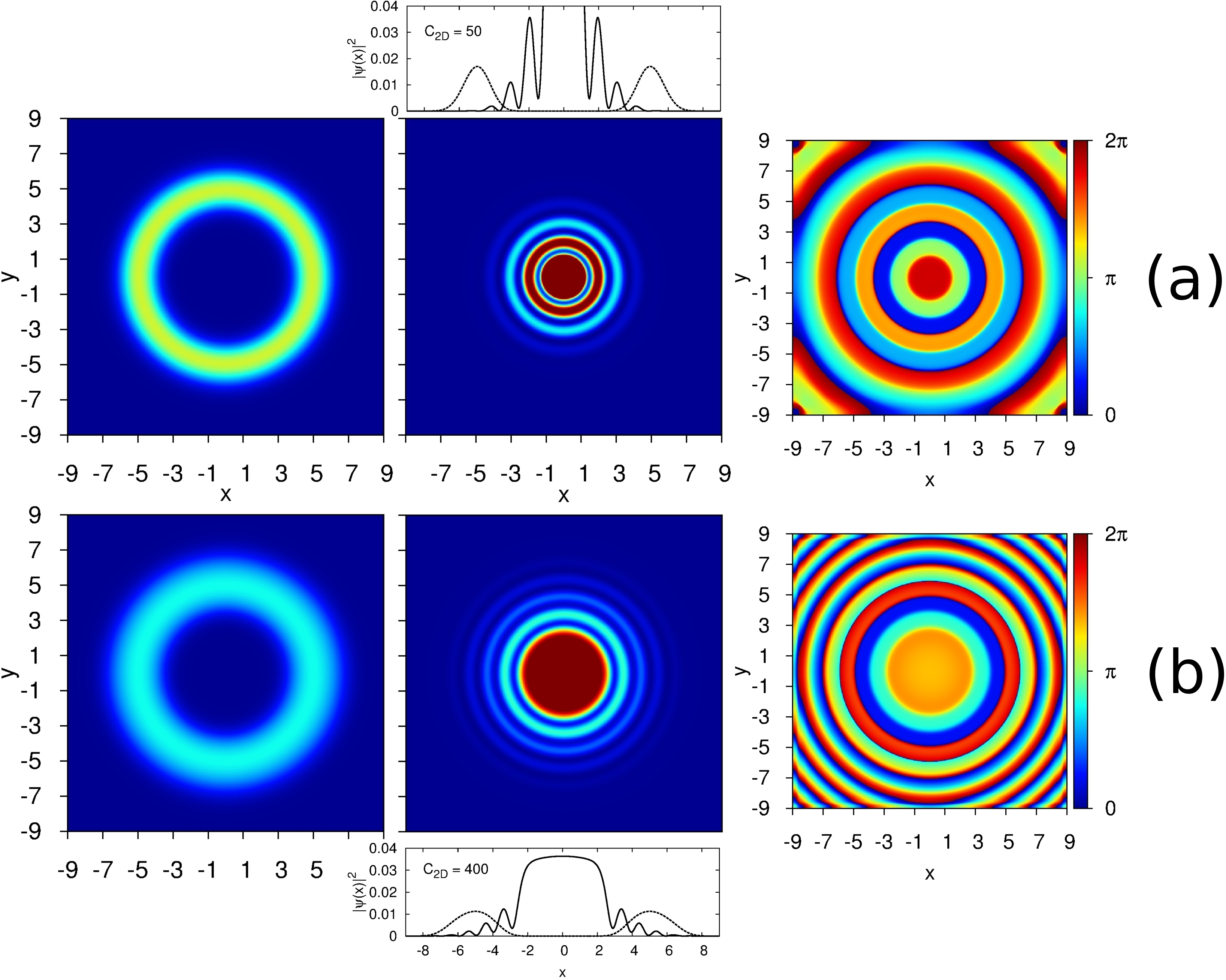}
\caption{\label{fig:MexHatFringes} Direct numerical simulation of the GPE~\eref{eqn:nls0} with $V_{\mathrm{trap}}$ given by~\Eref{eqn:V_MexicanHat}, and (a) $C_{\mathrm{2D}} = 50$ and (b) $C_{\mathrm{2D}} = 400$. The left panel shows the initial density distribution at $t = 0$, and the middle panel shows the same distribution at $t = 1.5$. The top and bottom insets show a slice along the $x$-axis through the cylindrically symmetric position distributions (solid line), and the initial ring is also shown (dashed line). The right panel shows the phase distributions corresponding to the densities of the middle panels. In both cases, the ring expands towards the centre, and once the opposite parts of the ring overlap, circular long-lived soliton-like fringes associated with phase steps appear in the position distribution. The central peak for the $C_{\mathrm{2D}} = 50$ case is 5 times higher than for $C_{\mathrm{2D}} = 400$, giving rise to a higher kinetic energy density. The colouring is the same in all figures.}
\end{figure}

The fringes of Fig.~\ref{fig:GaussApprox1} do not appear in the position distribution as long as the trapping potential forbids occupation in the central region. Still, these results suggest that if we start with a toroidal condensate at some finite radius and let the ring expand towards the origin, we should observe circular fringes when the opposite and in general different parts of the ring overlap with each other.

\subsection{\label{sec:sim}Self-interference of an expanding ring}

To gain preliminary understanding, we integrate the GPE~\eref{eqn:nls0} numerically in a time-dependent Mexican hat potential after the central region of the hat has been suddenly removed and the toroidal condensate allowed to expand freely towards the centre. The ground state is first found by propagating the GPE in imaginary time, and the propagation is done in real time. We use a split-operator Fourier method~\cite{0034-4885-58-4-001}. The potential we choose has the following form: 
\begin{equation}
\label{eqn:V_MexicanHat}
V_{\mathrm{trap}}(r,t) = \frac{1}{4} r^2 + 30 H(-t) \rme^{-0.1 r^2},
\end{equation}
where $H$ is the Heaviside step function. The simulation reveals that once the opposite parts of the ring start to overlap around the origin at $t \approx 1.5$, circular fringes develop around a bright central spot (see Fig.~\ref{fig:MexHatFringes}). This is the self-interference of the ring induced by the toroidal geometry.

We can also study more analytically the time-evolution of the wavefunction after the ring is allowed to expand inwards. We neglect the (weak) nonlinear interaction and also assume that the external potential vanishes after the toroidal trapping is released so that the time evolution is given by the free-particle linear Schr\"odinger equation.

To get the time evolution, we multiply the zeroth order Fourier-Bessel transform $\tilde{\varphi}_G(k)$ of $\varphi_G(r)$ by $\rme^{\rmi k^2 t}$ in $k$-space, and take the inverse transformation, viz.:
\begin{eqnarray}
\label{eqn:MH_an2}
\varphi_G(r,t) &= \int_{0}^{\infty} \mathrm{d} k\, k J_0 (kr) \tilde{\varphi}_G(k) \rme^{\rmi k^2 t} \\
&=\, \mathcal{N} \int_{0}^{\infty} \mathrm{d} k\, \mathrm{d} r' \, k r'  J_0 (kr) J_0 (kr') \rme^{-\frac{(r'-r_0)^2}{2\sigma^2}} \rme^{\rmi k^2 t}.
\end{eqnarray}
We note that while $\tilde{\varphi}_G(k)$ can be evaluated exactly similarly to~\Eref{eqn:GaussApprox1}, it is a good approximation to consider only the $m = 0$ term (see the Appendix):
\begin{equation}
\label{eqn:MH_an3}
\varphi_G(r,t) = \mathcal{N}\sigma^2 \int_{0}^{\infty} \mathrm{d} k\, k J_0 (kr) J_0(kr_0) \rme^{-\frac{k^2 \sigma^2}{2}} \rme^{\rmi k^2 t}.
\end{equation}

If $r_0 = 0$, then~\Eref{eqn:MH_an3} becomes exact and equivalent to~\Eref{eqn:MH_an2}, and can be evaluated in closed form to obtain an expanding radial Gaussian with no further dynamics;
\begin{equation}
\label{eqn:MH_an4}
\left. \varphi_G(r,t)\right|_{r_0 = 0} = \frac{\mathcal{N}\sigma^2}{\sigma^2-2\rmi t} \exp{\left( -\frac{r^2}{2\sigma^2 - 4\rmi t} \right)}.
\end{equation}
For $r_0 > 0$, on the other hand, the extra factor of $J_0(kr_0)$ will give rise to oscillations, i.e. the origin of the self-interference fringes of a ring condensate in a cylindrically symmetric system is a Bessel function as opposed to the typical cosine modulation that gives rise to interference in planar systems. Evaluating~\Eref{eqn:MH_an3} gives
\begin{eqnarray}
\label{eqn:MH_an5}
\fl \varphi_G(r,t) &= \mathcal{N}\sigma^2 \sum_{m = 0}^{\infty} \frac{2\left(-r_0^2\right)^{m}}{m!\left( 2\sigma^2 -4\rmi t\right)^{m+1}}\, _1F_1\left(1+m,1,-\frac{r^2}{2\sigma^2 - 4\rmi t} \right) \\
\label{eqn:MH_an5s}
\fl &=\, \frac{\mathcal{N}\sigma^2}{\sigma^2 - 2\rmi t} \exp\left(-\frac{r^2+r_0^2}{2\sigma^2 - 4\rmi t} \right) I_0\left( \frac{rr_0}{\sigma^2 - 2\rmi t} \right),
\end{eqnarray}
where the second equality follows after some algebra and has the form of a Skellam distribution for complex arguments. In the limit as $r_0 \to 0$,~\Eref{eqn:MH_an4} is a special case of~\Eref{eqn:MH_an5s}, which in general describes an expanding ring condensate of initial radius $r_0$.

We note that~\Eref{eqn:MH_an5s} is general given that $C_{\mathrm{2D}} = 0$, and it has the form of an exponential envelope modulated by $I_0$, the modified Bessel function of order 0. The self-interference fringes arise from $I_0$ as its argument is complex, but an exhaustive mapping of fringe periods, time scales, and contrasts as functions of $r_0$ and $t$ is beyond the scope of this work.
\begin{figure}
\centering
\subfigure{\includegraphics[width=0.45\textwidth] {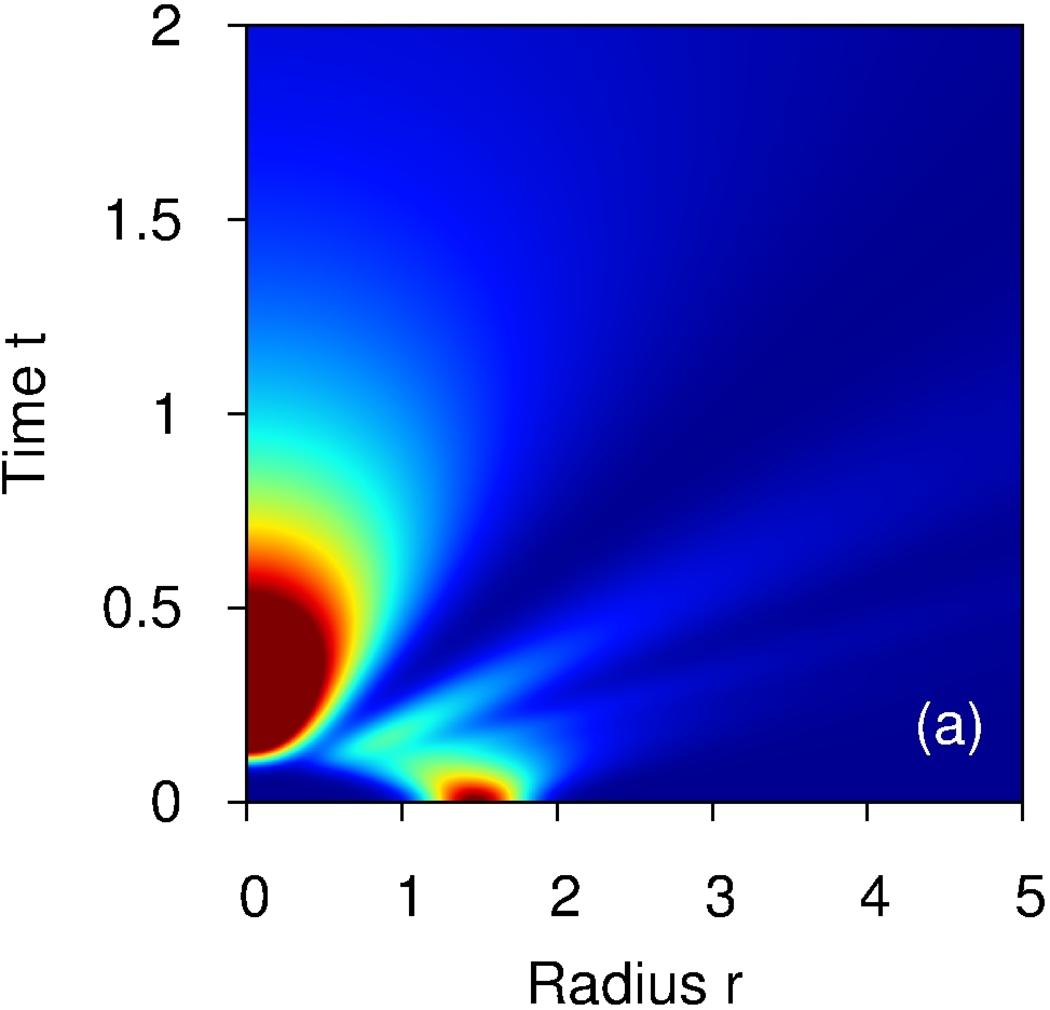}} 
\subfigure{\includegraphics[width=0.45\textwidth] {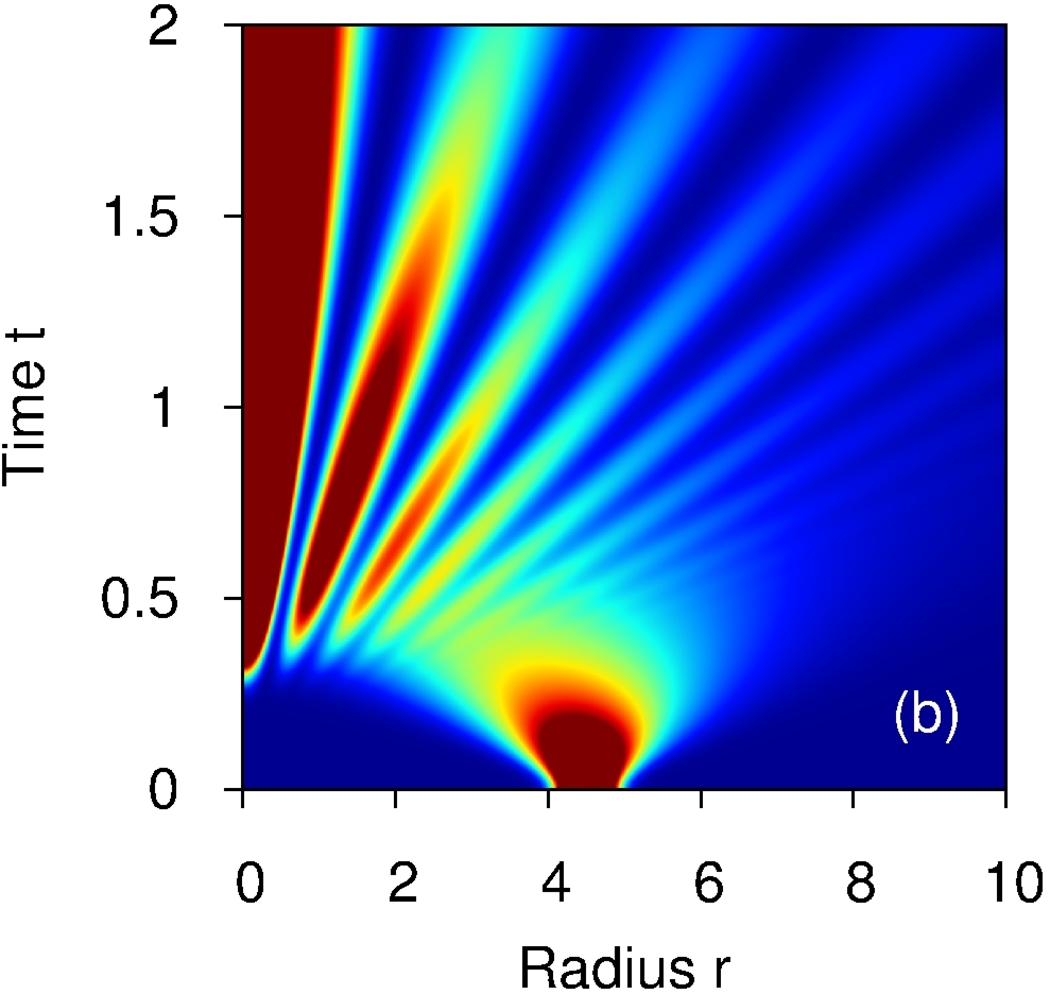}} 
\subfigure{\includegraphics[width=0.45\textwidth] {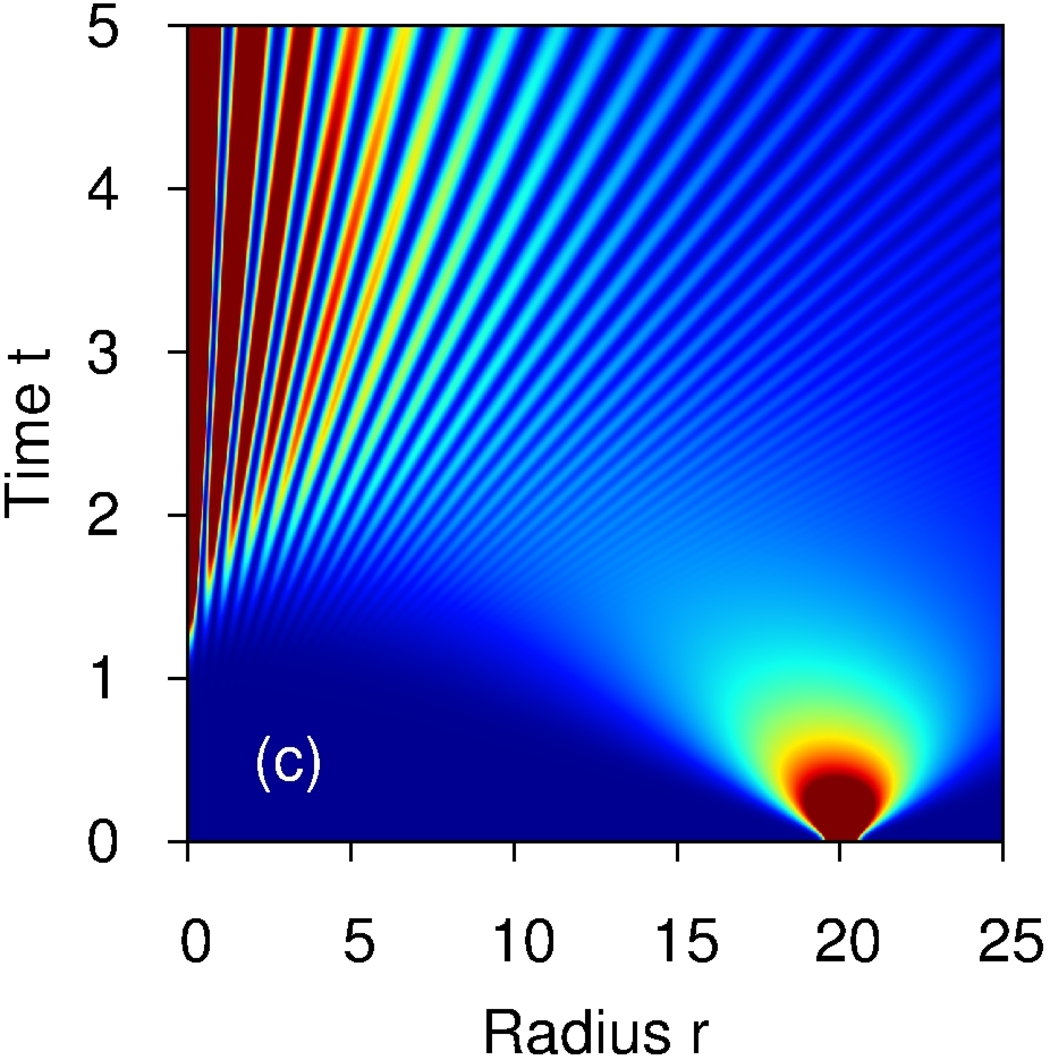}} 
\subfigure{\includegraphics[width=0.45\textwidth] {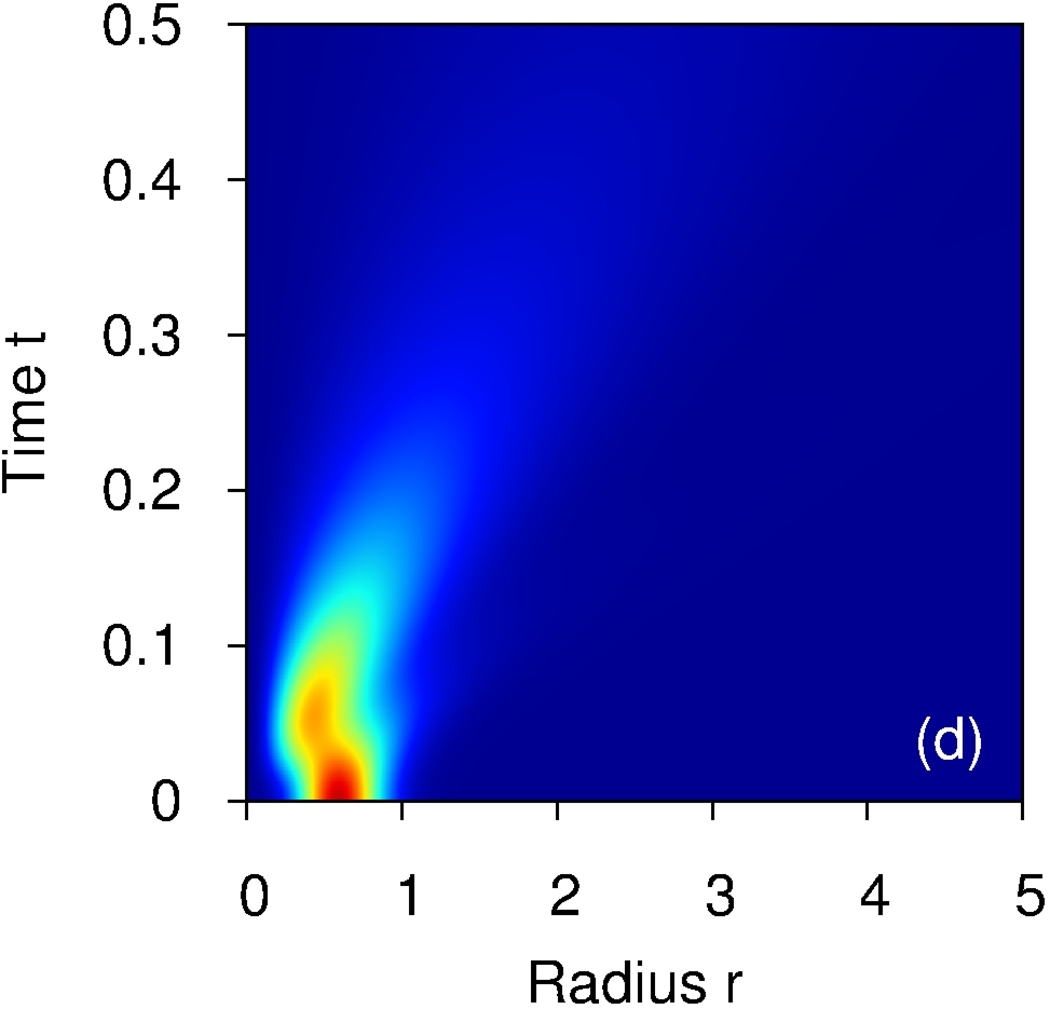}} \\
\caption{\label{fig:3} Density plot of a ring-shaped Gaussian condensate of initial radius (a) $r_0 = 1.5$, (b) $r_0 = 4.5$, and (c) $r_0 = 20$ expanding freely as a function of time [see~\Eref{eqn:MH_an5s}]. Self-interference fringes appear when opposite parts of the rings meet at $t \approx 0.1$, $t \approx 0.4$, and $t \approx 1.1$ respectively. Here $\sigma = \sqrt{2/20}$. (d) Density plot of a ring-shaped persistent current ($r_0 = 0.4$) expanding freely as a function of time [see~\Eref{eqn:vortex2s}]. There is no self-interference because the condensate cannot expand to $r = 0$ to overlap with itself. We have numerically confirmed that direct integration of the GPE with $C_{\mathrm{2D}} = 0$ gives the same results as the analytical evolutions shown here. The colouring is not to scale.}
\end{figure}
~\Fref{fig:3} (a-c) show the time evolution given by~\Eref{eqn:MH_an5s} for various $r_0$. All the essential features of Fig.~\ref{fig:MexHatFringes} are reproduced; the central bright peak and the circular self-interference fringes around it can be seen to emerge once the opposite parts of the ring overlap. 

As an interesting generalisation, we conclude this section by considering a vortex state (i.e. a persistent current) with winding number $\zeta$, given by $\varphi_v (r,\theta) = \varphi_G(r) r \rme^{\rmi\zeta \theta}$. Taking the $m = \zeta$ term in a similar fashion as above, but setting $\zeta = 1$ for definitess and assuming $r_0 \ll 1$ to avoid cumbersome expressions, we get
\begin{eqnarray}
\label{eqn:vortex2}
\fl \varphi_v(r,\theta,t) &= \mathcal{N}\sigma^4 \rme^{\rmi \theta} \int_0^\infty \mathrm{d} k\, k^2 J_1(kr) J_0(kr_0) \rme^{-\frac{1}{2}k^2\sigma^2} \rme^{\rmi k^2t} \\
\label{eqn:vortex2a}
\fl &= \mathcal{N}\sigma^{4} \rme^{\rmi\theta} r \sum_{m = 0}^{\infty} \frac{4(m+1)\left(-r_0^2\right)^{m}}{\, m! \left( 2\sigma^2 - 4 \rmi t\right)^{m+2}}\, _1F_1\left(2+m,2,-\frac{r^2}{2\sigma^2 - 4\rmi t} \right) \\
\label{eqn:vortex2s}
\fl &=\, \frac{\mathcal{N}\sigma^4 \rme^{\rmi\theta}}{(\sigma^2 - 2\rmi t)^2} \exp\left( -\frac{r^2+r_0^2}{2\sigma^2-4it} \right) \left[ rI_0\left( \frac{rr_0}{\sigma^2-2it} \right) - r_0I_1\left( \frac{rr_0}{\sigma^2-2it} \right) \right]
\end{eqnarray}
The vortex at $r = 0$ must be accompanied by a hole in the density to avoid a diverging kinetic energy. This makes the vortex state unstable in an unrotated simply-connected condensate~\cite{PhysRevLett.79.2164}, but on toroidal geometry, the vortex line can exist in the central forbidden region with the result that the persistent current state is stable. The presence of the phase singularity means that the opposite parts of the ring condensate cannot overlap in the free expansion, and hence their self-interference is not possible [see~\Fref{fig:3} (d)]. In~\cite{PhysRevA.86.013629}, it was experimentally observed that this central hole in the atomic density is present even after a long period of free expansion. 

\subsection{\label{sec:RDS}Experimental creation of ring dark solitons by toroidal self-interference}
\begin{figure}
\centering
\includegraphics[width=\textwidth] {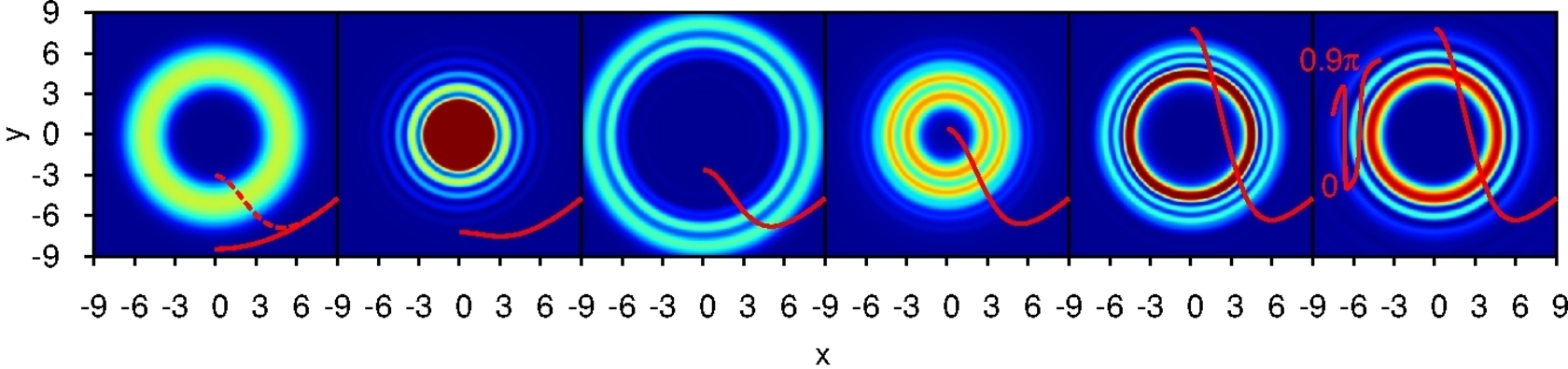}
\caption{\label{fig:cRDS} Density plots of the self-interference of the initial torus (first frame), and the subsequent generation of two ring dark solitons as given by numerically integrating the GPE~\eref{eqn:nls0} with $V_{\mathrm{trap}}$ given by~\Eref{eqn:V_RDS} and $C_{\mathrm{2D}} = 400$. The red solid line shows a slice of the potential along the positive $x$-axis. In the first frame, the dashed line shows the potential for negative $t$. In the last frame, also the phase steps of $\sim 0.9\pi$ over the RDSs are shown along the negative $x$-axis. From left to right, the times correspond to $t = 0, 1.5, 3.5, 4.8, 10.7,$ and $13.1$.}
\end{figure}
Earlier, we have proposed a protocol for the controlled creation of ring dark solitons (RDSs) by means of a time-dependent double-well trap~\cite{2013arXiv1309.0732T}. Ring dark solitons are a cylindrically symmetric extension of planar dark solitons, whose existence and stability depend on the nonlinearity and quasi-one-dimensionality of the system, respectively~\cite{PhysRevA.87.043601}. In~\cite{PhysRevA.74.043613}, RDSs are identified with the nodes of numerically found radial solutions of the GPE (with cylindrical symmetry) that approach the Bessel functions in the linear limit, which, as we have shown, also give rise to the self-interference fringes. To date, RDSs have not been experimentally observed in cold atomic quantum liquids, however, and here, we propose an alternative protocol that involves the use of the toroidal self-interference as a density imprinting mechanism for their creation. In other words, we let the BEC produce an interference pattern, which then evolves into ring dark solitons. We note that density imprinting has been experimentally demonstrated for the production of planar dark solitons~\cite{Shomroni09}, and that interference fringes have been shown in general to evolve into stable soliton-like structures in the case of two colliding condensates if the kinetic energy of the condensates during the collision does not dominate over the nonlinear self-energy~\cite{0953-4075-31-8-001}.

At least some of the fringes shown in~\Fref{fig:MexHatFringes} are soliton-like rings that are long-lived and oscillate back and forth in the harmonic trap. When they are at the turning points of their oscillation, the associated phase step has a magnitude close to $\pi$, which characterises dark solitons in general. As a proof-of-concept demonstration for the creation of ring dark solitons through toroidal self-interference, we consider the potential~\eref{eqn:V_MexicanHat}, but after the self-interference has taken place, we adiabatically ramp the central region back up. Because the condensate obtains kinetic energy upon the removal of the central region, we must replace it by a higher barrier to prevent the condensate from sloshing back and forth over it. As an example potential, we consider a linear ramp starting at $t = 1$ and lasting until $t = 8$:
\begin{equation}
\label{eqn:V_RDS}
\eqalign{
\fl V_{\mathrm{trap}}(r,t) = \frac{1}{4} r^2 + 30 \left[H(-t) + \right.
 \nonumber \cr
\left. 3H(t - 1)H(-t+8)\frac{t-1}{7} + 3H(t-8)\right] \rme^{-0.1 r^2} }
\end{equation}
with $C_{\mathrm{2D}} = 400$.

The time evolution given by the potential~\eref{eqn:V_RDS} above is shown in~\Fref{fig:cRDS}. Initially as the central region is removed, the condensate starts expanding and obtains kinetic energy. When the potential blocking the central region is later turned back on, the condensate is bound on the toroidal geometry again, albeit with some kinetic energy. As is evident in~\Fref{fig:cRDS}, in this case the self-interference prints two clear long-lived ring dark solitons. We have not observed the snake instability~\cite{PhysRevA.87.043601} in the time scales associated here.

Apart from the resulting motion of the condensate, we have demonstrated that it is possible to generate ring dark solitons in toroidal (and harmonic) traps by starting from a toroidal condensate that is let to interfere with itself. We note that such a protocol is easy to implement experimentally. For example, in~\cite{PhysRevLett.111.205301}, a condensate is prepared in a ring of radius $4\, \mu\mathrm{ m}$ using time-averaged painted potentials, comparable to the first frame in~\Fref{fig:cRDS}. The spatial resolution of the potential is stated to be $\sim 1.5\, \mu\mathrm{ m}$, which is more than enough to paint the central barrier in our case. As stated in~\Sref{sec:tb}, the physical scales corresponding to the experiment~\cite{PhysRevLett.111.205301} of our dimensionless units are $a_{\mathrm{osc}} = 0.8\, \mu\mathrm{ m}$, $C_{\mathrm{2D}} \sim 200-2000$, and one unit of time is $\sim 1.5$ ms. Our choice of $C_{\mathrm{2D}}$ in~\Fref{fig:cRDS} falls within this regime.

\section{Conclusions}

In conclusion, we have studied the self-interference of a toroidal condensate both with and without a persistent current. We investigated the Wigner function representation of the ring condensate to see how the self-interference arises from the toroidal geometry. In particular, for a ring of zero radius, all interference phenomena vanished. We predicted both numerically and analytically the appearance of circular fringes if a toroidal condensate is allowed to freely expand towards the origin. In contrast, in the presence of a toroidal persistent current, we showed that there cannot be any ring-induced self-interference because the opposite parts of the condensate cannot overlap in the free expansion. Furthermore, we gave a proof-of-principle demonstration of how ring dark solitons can be experimentally created using the toroidal self-interference. 

The results are important because they open new possibilities for demonstrating the quantum wave-nature of matter on toroidal geometry. This enables probing the phase-coherence over the ring, for example, and also to experimentally create ring dark solitons in ring traps. Letting the ring dark solitons decay into necklaces consisting of vortex-antivortex pairs leads eventually to quantum turbulence, but before that they have been shown to recombine back into the ring dark solitons~\cite{PhysRevA.87.043601, 2013arXiv1309.0732T}. Experimental study of the coherent vortex dynamics before the onset of quantum turbulence forms another interesting area for future work.

\ack
We acknowledge the support of the Academy of Finland (grant 133682) and Jenny and Antti Wihuri Foundation. We thank Kalle-Antti Suominen for fruitful discussions, and Vladimir S. Ivanov for obtaining the simplified Eqs.~\eref{eqn:MH_an5s} and~\eref{eqn:vortex2s}.

\section*{References}
\bibliographystyle{unsrt}
\bibliography{../references}

\appendix
\section{\label{ap:1}Evaluation of~\Eref{eqn:GaussApprox1}}
~\Eref{eqn:GaussApprox1},
\begin{equation}
\label{eqn:GaussApprox1aa}
W_{\varphi_G}(0,p) = \mathcal{N}^2 \int_{0}^{\infty} \mathrm{d} u\, u p J_0 (up) e^{-\frac{(u/2-r_0)^2}{\sigma^2}},
\end{equation}
is a non-standard integral involving the zeroth order Bessel function $J_0$. Let us first make the substitution $t = u/2 - r_0$ in~\Eref{eqn:GaussApprox1aa}: 
\begin{equation}
\label{eqn:GaussApprox2}
W_{\varphi_G}(0,p) \approx \mathcal{N}^2 \int_{0}^{\infty} \mathrm{d} t\, 4(t+r_0) p J_0 [2p(t+r_0)] e^{-\frac{t^2}{\sigma^2}},
\end{equation}
where we have assumed that $r_0$ is small enough that the lower limit of the integral can be set to $0$. The error thus introduced could be corrected by numerically evaluating the integral in~\Eref{eqn:GaussApprox2} from $t = -r_0$ to $t = 0$. Let us then make use of the Bessel function addition theorem
\begin{equation}
\label{eqn:BesselAddition}
J_n(y + z) = \sum_{m = -\infty}^{\infty} J_m(y)J_{n-m}(z)
\end{equation}
together with the symmetry property $J_{-m}(z) = (-1)^m J_m(z)$ to obtain
\begin{equation}
\label{eqn:GaussApprox3}
\eqalign{
\fl W_{\varphi_G}(0,p) = \sum_{m = 1}^{\infty} 8\mathcal{N}^2 p (-1)^m J_m(2pr_0) \int_{0}^{\infty} \mathrm{d} t\, (t+r_0) J_m (2pt) e^{-\frac{t^2}{\sigma^2}}
 \nonumber \cr
+\, 4\mathcal{N}^2 p J_{0}(2pr_0)\int_{0}^{\infty} \mathrm{d} t\, (t+r_0) J_0 (2pt) e^{-\frac{t^2}{\sigma^2}}.}
\end{equation}
After evaluating the integrals in~\Eref{eqn:GaussApprox3}, we obtain
\begin{equation}
\label{eqn:GaussApprox4}
\eqalign{
\fl W_{\varphi_G}(0,p) = \sum_{m = 1}^{\infty} 4\mathcal{N}^2 (-1)^m J_m(2pr_0) 
 \nonumber \cr
\times p^{m+1} \sigma^{2+m} \left[\Gamma \left(1 + \frac{m}{2}\right) \frac{_1 F_1 (1+\frac{m}{2},1+m,-p^2 \sigma^2)}{\Gamma(1+m)} \right.
 \nonumber \cr
 \qquad \qquad + \left. \frac{r_0}{\sigma}\Gamma \left(\frac{1+m}{2} \right) \frac{_1 F_1 (\frac{1+m}{2},1+m,-p^2 \sigma^2)}{\Gamma(1+m)}  \right]
  \nonumber \cr
 +\, 2 \mathcal{N}^2 p J_{0}(2pr_0)\sigma^2 e^{-p^2\sigma^2}\left[1 + \frac{\sqrt{\pi}r_0 I_0(p^2 \sigma^2 / 2) }{\sigma} e^{\frac{p^2\sigma^2}{2}} \right],}
\end{equation}
where $I_0$ is the modified Bessel function of order 0, $\Gamma$ is the Gamma function, and $_1 F_1$ is the confluent hypergeometric function. If $r_0 = 0$,~\Eref{eqn:GaussApprox4} reduces to $2\mathcal{N}^2 p\sigma^2 e^{-p^2 \sigma^2}$, as required. In the right panel of Fig.~\ref{fig:GaussApprox1}, we have cut the infinite summation over $m$ with $m_{\mathrm{max}} = 10$.

\end{document}